# Manipulation of graphene's dynamic ripples by local harmonic out-of-plane excitation[1]

A. Smolyanitsky* and V.K. Tewary

Materials Reliability Division, National Institute of Standards and Technology, Boulder, CO 80305

*corresponding author: alex.smolyanitsky@nist.gov

**Abstract**

With use of carefully designed molecular dynamics simulations, we demonstrate tuning of dynamic ripples in free-standing graphene by applying a local out-of-plane sinusoidal excitation. Depending on the boundary conditions and external modulation, we show control of the local dynamic morphology, including flattening and stable rippling patterns. The nonlinear lattice response observed at higher excitation amplitudes also suggests a possibility of exciting transverse solitons in free-standing graphene. In addition to studying the dynamic response of atomically thin layers to external time-varying excitation, our results open intriguing possibilities for modulating their properties via local dynamic morphology control.

Atomically thin layers present a unique case in solid-state physics, where two-dimensional confinement results in remarkable electronic properties of graphene[1-4], the first truly two-dimensional (2-D) crystal[1,5]. The effect of local landscape of the layer on its properties is important, because free-standing graphene is naturally rippled[6]. These ripples have been attributed to a dynamic effect of finite temperature,[7] ultimately resulting in the destruction of the long-range order in 2-D crystals, in qualitative agreement with the long-standing Mermin-Wagner theorem[8] and the statistical considerations of the thermally fluctuating membranes[9, 10]. This mechanism is closely related to the fundamental instability of the Bose-Einstein statistics in strict 2-D[11] and can be viewed as graphene's natural way of existing in three-dimensional space. Static rippling has been suggested as an important additional mechanism affecting the local morphology, arising from disrupted bonding coordination at the edges, as well as from lattice defects dispersed throughout the layers[12, 13]. Static wrinkles in substrate-bound graphene have also been reported, along with their effect on graphene's electronic properties.[14]

---

[1]Contribution of the National Institute of Standards and Technology, an agency of the US government. Not subject to copyright in the USA.

Control of the texture in suspended graphene sheets by local strain[15], due to rolling[16] and dynamic mechanical excitation of graphene[17] were previously reported. These findings are especially important, because recent experimental work revealed local quantum confinement in graphene drumheads[18], which results from adjustable strain-induced pseudomagnetic fields previously predicted theoretically[19]. In light of these reports, our ability to control the local morphology of suspended atomically thin layers may result in a new class of mechanically controlled quantum dots.

Manipulation of the local dynamic shape of suspended atomic layers can be achieved via dynamic time-varying excitation, similar to the case of vibrating drumheads. Here, we report the results of molecular dynamics (MD) simulations of an out-of-plane sinusoidal excitation locally applied to a thermally fluctuating graphene membrane. We demonstrate that the coexistence of stochastic (thermal) ripples and the externally excited out-of-plane vibrations in an atomically thin membrane subject to energy dissipation results in a number of interesting phenomena, including significant overall flattening and emergence of stable dynamic ripple patterns. We show high controllability of the wavelength and amplitude of the dynamic membrane ripples, depending on the excitation amplitude and frequency, beyond the previous results for electromechanical graphene resonators[17].

Results and Discussion

The main quantity of interest in our simulations was the mean-square displacement (MSD) of the atoms in the out-of-plane direction (the membrane initially positioned in XY-plane at Z = 0), defined as

$$\langle h_{tot}^2 \rangle = \langle \frac{1}{N} \sum_N z_i^2 \rangle, \qquad (1)$$

where $z_i$ is the out-of-plane (Z-) coordinate of the $i$-th atom in the system, $N$ is the total number of atoms, and $\langle . \rangle$ represents the time-average. Without external excitation, the amount of stochastic out-of-plane rippling $\langle h_0^2 \rangle$ is directly proportional to the equilibrium temperature, as shown in the inset of Fig. 1, consistent with previous results [7,9]. At 300 K, we calculated $\bar{h}_0 = \sqrt{\langle h_0^2 \rangle} = 0.27$ Å, which corresponds to an average height of $\sqrt{2}\,\bar{h}_0 = 0.38$ Å, about an order of magnitude higher than that obtained from the Green's function based analysis of vibrations *at the atomic site*[20] and ~1.8 times lower than for a sample of similar size obtained from Monte-Carlo simulations[7]. Our thermal ripple height is also significantly lower than the ~1 nm obtained in a continuum model of externally excited graphene membranes.[21] One must keep in mind that the average ripple height scales with the membrane size, as well as any external input and temperature[9] and thus a direct comparison is only reasonable for heights properly normalized with respect to these factors. Within the selected temperature range, there is no indication of significant anharmonic effects, in agreement with [7]. The external excitation was a sinusoidal force $F(t) = F_0 \cos(\omega t)$ applied to a single atom at the intersection of the diagonals of the membrane, as shown in Fig. 1. Such input caused the induced vibrations to spread throughout the entire membrane[22] (also see Section 1 of the Supplementary Information).

Shown in Fig. 2 (a) are the results for the root-mean-square of the out-of-plane displacements due to an external force normalized to the corresponding value without external excitation $\varepsilon \equiv \sqrt{\langle h_{tot}^2 \rangle / \langle h_0^2 \rangle}$. The forcing frequency $f = \omega/2\pi$ was swept over the indicated range at four different values of $F_0$, as shown. For $F_0 = 8\ nN$ and $f < 5\ THz$, $\varepsilon$ is considerably higher than unity and dynamic rippling patterns emerge, as shown in Fig. 2 (b) for $f = 1\ THz$, while for higher excitation frequencies $\varepsilon < 1$, and an intriguing effect of flattening takes place, as also shown in Fig. 2 (b) for $f = 10\ THz$. The two particular local minima in $\varepsilon$ are at $f \sim 11\ THz$ and $f \sim 22\ THz$. Toward the frequency of $40\ THz$, the membrane appears to lose all its sensitivity to external excitation and $\varepsilon$ becomes close to unity, while the ripple distribution of the sheet becomes nearly identical to the case without external excitation (Fig. 2 (b)). For $F_0 = 4\ nN$, $2\ nN$, and $1\ nN$ the data in Fig. 2 (a) show nearly identical behavior, although the positions of the minima exhibit shifting to the left. This shifting is likely due to the decreasing anharmonicity in the response with decreasing $F_0$. The overall effect on $\varepsilon$ also decreases with decreasing $F_0$, as one may expect. We shall come back to the discussion of the observed dynamic patterns at low frequencies and their significance later in the text. We now consider the observed flattening effect and demonstrate that it is a result of local dissipation, which can be explained from the basic standpoint of thermodynamics.

Let us consider the energies associated with the atomic out-of-plane motions in a two-dimensional thermally fluctuating membrane. In absence of external excitation, the average kinetic energy of the atoms $E_0$ (subject to dissipation both in nature and in the simulation) is determined by the effective average temperature of the sheet, calculated as the ensemble-average of the per-atom kinetic energy. At the same time, the out-of-plane MSD defined by Eq. (1) in absence of external excitation, $\langle h_0^2 \rangle$ is also proportional to the temperature[9, 10] due to equality between average potential and kinetic energies in the oscillatory steady-state:

$$T_0 \propto E_0 = A\langle h_0^2 \rangle, \qquad (2)$$

where $A$ is a constant, which depends on the out-of-plane bending rigidity and size of the membrane, arising from the statistical analysis of the thermal ripples[9]. When an external excitation is applied, we assume a harmonic response, *i.e.* any out-of-plane stochastic (thermal) ripples with MSD of $\langle h^2 \rangle$ coexist with the deterministic externally induced ripples with MSD of $\langle h_{ext}^2 \rangle$ (the brackets here represent a time-average), and $\langle h, h_{ext} \rangle = 0$. As a result of external excitation, as well as any energy loss, the new overall temperature is $T$; the overall energy is a sum of the thermal energy $A\langle h^2 \rangle$ and an additional component $E_{ext}$ carried by the externally induced ripples:

$$T \propto E = A\langle h^2 \rangle + E_{ext}. \qquad (3)$$

It is paramount to keep in mind that Eq. (3) is merely a sum of average kinetic energies, where $A\langle h^2 \rangle$ is the component incorporating all stochastic contributions expressed for convenience as a function of $\langle h^2 \rangle$ and $E_{ext}$ is the externally excited kinetic energy contribution. Their distinction is only possible within the harmonic approximation. As we show later, the result we seek does not explicitly depend on the constant $A$, provided it is identical in Eqs. (2) and (3) (guaranteed by the assumption of harmonic response).

Introducing $\lambda$ as the general factor of temperature increase as a result of external excitation, as well as any energy loss ($\lambda \equiv T/T_0$) and combining Eqs. (2) and (3), we obtain:

$$\langle h^2 \rangle = \lambda \langle h_0^2 \rangle - \frac{E_{ext}}{A}. \tag{4}$$

Once again, we note that $\langle h^2 \rangle$ in Eqs. (3) and (4) is only the stochastic (thermal) portion of the overall out-of-plane vibrations, which cannot be obtained from the simulations. The total amount of simulated rippling defined by Eq. (1) includes the externally induced contribution $\langle h_{ext}^2 \rangle$:

$$\langle h_{tot}^2 \rangle = \langle h^2 \rangle + \langle h_{ext}^2 \rangle = \lambda \langle h_0^2 \rangle - \frac{E_{ext}}{A} + \langle h_{ext}^2 \rangle. \tag{5}$$

From Eq. (5), recalling Eq. (2) and the equipartition theorem[23], we can express the out-of-plane rippling ratio between the cases with and without external excitation, obtained from simulations. Here, we define it as the ratio of the square roots of the corresponding MSD values, or a ratio of the RMSDs:

$$\varepsilon \equiv \sqrt{\frac{\langle h_{tot}^2 \rangle}{\langle h_0^2 \rangle}} = \sqrt{\lambda + \frac{\langle h_{ext}^2 \rangle}{\langle h_0^2 \rangle} - 2\frac{E_{ext}}{k_b T_0}}, \tag{6}$$

where $k_b$ is the Boltzmann constant. Equation (6) demonstrates the effect of external excitation on the overall amount of dynamic rippling in the form of a competition between the magnitude of the externally induced ripples and the amount of kinetic energy they carry.

As a general check for Eq. (6), it is evident that if the energy is conserved by the membrane and $\lambda = \left(1 + 2\frac{E_{ext}}{k_b T_0}\right)$, the effect of flattening cannot be observed, because $\varepsilon = \sqrt{1 + \frac{\langle h_{ext}^2 \rangle}{\langle h_0^2 \rangle}} > 1$, guaranteed for a non-zero excitation amplitude, as expected. Note that flattening may still be possible even if the energy is conserved. With strong anharmonicity, $\langle h, h_{ext} \rangle \neq 0$ must be present in Eq. (3). In addition, the values of $A$ may no longer be identical between Eqs. (2) and (3) and thus flattening may occur as a result of anharmonic stabilization of the lattice long-range order[7,9].

For a harmonic system we are considering, however, flattening is possible only in a non-adiabatic case, when $1 < \lambda < \left(1 + 2\frac{E_{ext}}{k_b T_0}\right)$, according to Eq. (6). The value of $\lambda$ is not freely adjusted to fit the simulation data, its value is obtained directly from the simulations. It is important to keep in mind that $\lambda$ depends on the external excitation, as well as the membrane's thermal conductivity (i.e. graphene-specific) and the dissipative properties of the heat sink represented by the thermostat. Ultimately, the externally induced portion $\langle h_{ext}^2 \rangle$ of the overall vibrations must be obtained in terms of the source frequency $\omega$ and the amplitude $F_0$; the external contribution to the kinetic energy is then automatically obtained, because $E_{ext} \propto \langle h_{ext}^2 \rangle \omega^2$.

We obtained the following closed-form approximation for the MSD of the externally excited ripples:

$$\langle h_{ext}^2 \rangle \cong \frac{1}{4} \frac{F_0^2}{m^2(\gamma^2 \omega^2 + (\omega^2 - \omega_0^2)^2)}, \qquad (7)$$

where $\gamma$ is effective damping in graphene as a result of energy loss in the membrane, $f_0 = \omega_0/2\pi = 0.29\ THz$ is the resonant response frequency of the graphene membrane with dimensions as stated earlier, and $m$ is the mass of the carbon atom. Above, $\omega$ is the frequency of external excitation. The kinetic energy carried by these ripples is $E_{ext} = \frac{m}{2} \langle h_{ext}^2 \rangle \omega^2$. The derivation and the approximations made are described in Section 1 of the Supplementary Information. Note that $\langle h_{ext}^2 \rangle$ and $E_{ext}$ oscillate at $2\omega$. Here, $\gamma$ is a fitting parameter, because no direct thermostatting was applied to and thus no effective viscosity was preset for the simulated sample, as described earlier. Also, regardless of the boundary conditions (periodic or clamped), the presence of harmonic functions results in an expression similar to Eq. (7). Even if the membrane shape is different, the expression of interest is identical to Eq. (7), except for the pre-factor on the right different from $1/4$.

Note that the frequency regions of interest (see Fig. 2 (a)) decrease with the effective sample size as $1/L$ (see Section 1 of the Supplementary Information). Thus, compared with the nanometer-sized samples, these regions are expected to shift toward giga- and megahertz for micron- and millimeter-sized samples, respectively, consistent with experiment[17]. The ripple magnitudes (both thermal and especially externally induced – see Eq. 7) are also expected to increase with the increased sample size in the corresponding frequency regions of interest. Finally, one must also realize that the electronic contribution to the dynamic response to the excitation is not included in our simulations. Clearly, this contribution can be significant, especially at high excitation frequencies. We therefore believe that our results are qualitatively closer to those one may expect at significantly lower frequencies of excitation, applied to larger samples.

Shown in Fig. 3 (a) is the normalized simulated RMSD of the out-of-plane displacement $\varepsilon = \sqrt{\frac{\langle h_{tot}^2 \rangle}{\langle h_0^2 \rangle}}$ at $F_0 = 4\ nN$ versus frequency from Fig. 2 (a), alongside that obtained from Eqs. (6) and (7). Each point in the analytical curve was obtained by using the simulated temperature shown in Fig. 3 (a). In all cases, we found $\lambda = T/T_0 < \left(1 + 2\frac{E_{ext}}{k_b T_0}\right)$. There are some general differences between the simulated and the analytically calculated data, including the positions of the local minima. These differences are attributed mainly to the relative simplicity of our analytical model, compared to the atomistic simulation. Specifically, the response of the membrane is represented by the single (1,1) mode (see Section 1 of Supplementary Information) and no anharmonic effects are taken into account. Finally, no discrete atomistic effects (especially important for the highest values of $\omega$ and thus short wavelengths) are taken into account. However, we see that flattening ($\varepsilon < 1$) is confirmed by the analytical model, and the overall trends are qualitatively similar in the frequency range where flattening occurs.

The underlying cause for the observed flattening is revealed by examining Eq. (6), given Eq. (7) and strongly bearing in mind that $E_{ext} \propto h_{ext}^2 \omega^2$. Three main regimes of response to excitation are therefore identified: i) low $\omega$ and high $h_{ext}^2$, ii) intermediate $\omega$ and $h_{ext}^2$, and iii) high $\omega$ and very low $h_{ext}^2$. It is useful to consider these cases in the context of the spectra in Figs. 3 (b, c), where we show the Fourier transforms of the simulated $h_{tot}^2(t) = \frac{1}{N}\sum_N (z_i(t))^2$ with and without excitation. In case (i), when a low-frequency excitation is applied, the large-amplitude out-of-plane vibrations carry a small kinetic energy contribution $E_{ext}$, and thus the excited membrane is subject to almost no extra energy loss. At the same time, a large induced peak $h_{ext}^2$ is added to the spectrum at twice the excitation frequency, as shown in Fig. 3 (c) for the excitation at 2.5 THz. As a result, the overall amount of rippling increases significantly, compared to the case without external excitation. In case (ii) the dissipative effect plays the major role, because the externally contributed kinetic energy $E_{ext}$ is now quite high, while $h_{ext}^2$ is moderate. As a result, the slow thermal vibrations at frequencies below *~7.5 THz* mostly responsible for the ripples are effectively lowered due to dissipation. At the same time, a moderate externally induced $h_{ext}^2$ peak is added to the spectrum. This "spectral intrusion" can be readily observed in Figs. 3 (b, c) for the excitation at 10 THz. Finally, in case (iii) of high excitation frequencies both $h_{ext}^2$ and $E_{ext}$ contribute negligible amounts, because they decrease as $1/\omega^4$ and $1/\omega^2$, respectively (see the spectra of $h_{ext}^2(t)$ in Fig. S1 in the Supplementary Information). Note the large anharmonic peaks present for $F_0 = 8\ nN$, decreasing at lower excitation amplitudes in Fig. 3 (b). This observation supports our previous remarks and agrees with the general discussion of anharmonicity in thermally fluctuating membranes [9]. It is noteworthy that the Eq. (6) is limited in determining the value of $\varepsilon$ as a function of varying temperature. This limitation is mainly due to $\langle h_{ext}^2 \rangle$ given by Eq. (7), which is a function of the effective viscoelastic damping $\gamma$, expected to depend on $T_0$. As a result, although we were able to use Eq. (6) to establish a general qualitative picture in Fig. 3 (a) at $T_0 = 300$ K, the temperature dependence of $\varepsilon$ may suffer, demonstrating the estimative nature of our analytical model. Consider, for instance, the results in Fig. 4 (a) and (b), where we plot $\varepsilon(T_0)$ obtained from simulation and from using Eq. (6), respectively. We used a constant excitation force of $F_0 = 2\ nN$ and plotted the temperature dependence for the excitation frequency of 1, 5, and 10 THz. A constant value of $\gamma$ was used, identical to that used to obtain the results shown in Fig. 3(a). We can see that qualitative agreement between the simulation and Eq. (6) only exists at f = 1 THz, when, as explained earlier, $\langle h_{ext}^2 \rangle$ is large, while $E_{ext} \propto h_{ext}^2 \omega^2$ is negligible and thus causes little dissipation (analytically, the dissipative term in the denominator of Eq. (7) is small). As a result, Eq. (6) for low excitation frequency is approximated as $\varepsilon \cong \sqrt{\lambda + \frac{\langle h_{ext}^2 \rangle}{\langle h_0^2 \rangle}}$ and with $\langle h_0^2 \rangle \propto T_0$, we expect a decreasing trend with increasing temperature, reproduced in both Fig. 4 (a) and (b). The situation changes when the excitation frequency increases: both $E_{ext}$ and the damping-dependent term in Eq. (7) increase as $\omega^2$. Not surprisingly, the agreement between simulated data and Eq. (6) becomes poor. Even if we assume ideal dissipation in all cases ($\lambda = 1$), the apparent discrepancies observed can still be attributed to the effective viscosity assumed constant and independent of temperature. Although obtaining the $\gamma(T_0)$ dependence is beyond the scope of this work and the preceding discussion is presented to demonstrate the limitations of Eq. (6), it also highlights the necessity of finding out graphene's effective viscosity as a function of temperature.

Perhaps even more intriguing than the effect of flattening is the case of low excitation frequency mentioned above. Shown in Fig. 4 (a) are the frequency-dependent patterns obtained for the rectangular membrane. To further quantify these patterns, we calculated the reciprocal space spectra of $h_{tot}^2(x, y)$, similar to the frequency domain data presented in Figs. 3 (b, c). Shown in Fig. 5 (b) are the spectra normalized with respect to $\langle h_{tot}^2 \rangle$ to bring out the relative strength of the externally excited ripples, compared to the overall amount of rippling. The curves in Fig. 5 (b) have also been artificially shifted relative to each other along the Y-axis for further clarity. The observed peaks correspond to *half the wavelength* of the ripples (due to the square of the quantity of interest) and have been carefully compared with the ripple spacings, as calculated from the membrane structure snapshots in real space. The inset in Fig. 5 (b) shows the ripple wavelength $\lambda_{ripple}$ as a function of excitation frequency. The dashed line in the inset is a rough fit using a simple $\lambda_{ripple} = c_{eff}/f$ relationship for an interference pattern, which is what is essentially observed, despite the thermal smearing. Here, $c_{eff} = 5207 \frac{m}{s}$ is the effective speed of sound, reasonably close to $6000 \frac{m}{s}$ we used in our previous calculations (see section 1 of the Supplementary Information). We must note that a perfect interference pattern is only possible with ideal boundaries (ideally transparent in our case; also valid for an ideally reflecting boundary). The reality, however, is that the boundary is partially reflecting and transmitting due to the thermostat in the simulation. In an experiment, the same effect is expected from the energy loss due to substrate, depending both on the substrate itself and graphene-substrate interface properties. In addition, one can see that the externally excited peaks have significant width and thus represent a considerable departure from harmonic response, as also mentioned earlier in the discussion of the vibrational spectra in the frequency domain.

The effect of changing temperature goes beyond the simple thermal smearing of the externally driven ripples. Shown in Fig. 5 (c) is a family of normalized spatial spectra obtained for various temperatures. The relative "drowning" of the externally excited ripples in the thermal oscillations is well expected and is clearly observed as the temperature increases, with externally excited ripples significantly smeared at 400 K. However, the peaks associated with the externally excited ripples shift and split with temperature without a clearly defined trend, as shown in Fig. 5 (c). The effective ripple wavelength range is 2 to 4 nm in response to an identical external excitation and is a result of temperature changes. This behavior cannot be explained from the standpoint of harmonic response and, in addition to the boundary losses (which change with temperature, as discussed earlier), is likely due to significant anharmonic coupling between thermal and driven ripples. A precise analytical relationship between temperature, the strength of anharmonic coupling and its overall effect on the spatial distribution of the externally excited ripples presents solid ground for further work. For additional data on the spatial distribution of the ripples and their short-wavelength behavior from the normal-normal correlations, also see Section 4 of the Supplementary Information.

Note that these controllable patterns appear for both periodic and reflective finite boundaries (see Fig. S2 in the Supplementary Information). Also, we should note that the only requirement for obtaining these rippling patterns is that the region where the external force is applied is significantly smaller than the characteristic size of the membrane. The effective size of the excitation region will limit the shortest excitable ripple wavelength and may cause additional anharmonic effects (also see section 5 of Supplementary Information).

In the case of atomically thin layers, surface morphology directly affects the electronic properties. Therefore, there exists a possibility for transverse modulation in terms of the reported dynamic patterns. Although the presence of a thermal component makes these patterns less regular than in an ideal interference pattern, they do preserve periodicity and amplitude, depending only on the source frequency. Such a possibility is extremely attractive, given the recent report of experimentally observed quantum confinement as a result of local strain-induced pseudomagnetic fields[18]. Because these fields are present both at the boundary and throughout the locally suspended layer[19], one could probe the effect of dynamic patterns on the quantum Hall effect in the suspended graphene sheets, depending on the external modulation. Moreover, as supported by the observation of strong nonlinear peaks in Fig. 3(b), generation of lattice solitons[24], or nearly isolated lattice vibration packets should be possible. In addition to the possibility of dynamically controlled quantum dots via patterned lattice distortions, optical response of the material may also be dynamically tunable. The local morphology of free-standing layers has a significant effect on the plasmon response of these materials[25-28] and therefore the described dynamic patterns may be utilized for control of their optical properties.

**Conclusions**

In summary, we described a method of dynamic control over the morphology of free-standing graphene membranes by applying a local sinusoidal out-of-plane excitation. Effects from overall flattening to stable dynamical ripple patterns, depending on the excitation amplitude and frequency, were presented. The described excitation could probably be applied optically (with a pulsed laser), mechanically, or by applying a local time-varying magnetic field to a current-carrying graphene sample. The use of such local morphology control may result in an effective modulation of the electrical and optical properties of atomically thin layers. Although our work was focused on graphene, the results are generally applicable to atomically thin layers in terms of using a time-varying force for studying their dynamic response and modulating their properties.

**Methods**

The suspended graphene membranes (dimensions 13.6 nm × 15.8 nm, 8192 atoms) were positioned in the XY-plane and set up as shown in Fig. 1. In-plane periodic boundaries were imposed to minimize the effect of wave reflections. The equations of motion were integrated in three dimensions with a time step of 1 fs. The interatomic interactions were defined by the Tersoff-Brenner bond-order potential, which is well-established in describing a variety of static and dynamic lattice properties of graphene and carbon nanotubes[12, 29-32].

Because the thermal fluctuations may be influenced by the simulated heat bath the system is immersed into, the temperature control in the MD simulations was applied in an extremely careful way. The widely used Nosé-Hoover thermostat[33] was applied to a 1-nm wide region at the perimeter of the membrane, representing the heat sink effect of the sample mount, as denoted by "thermostatted atoms" in Fig. 1, to ensure that no preset amount of effective viscoelasticity was imposed upon the entire membrane. All averages of interest were calculated for the non-thermostatted "free atoms" (Fig. 1). Those atoms were allowed to move freely, governed only by the interatomic interactions. The velocities at the start of all simulations were generated according the Maxwell-Boltzmann distribution at $T_0 = 300$ K and the temperature maintained at the perimeter with and without external excitation was also 300 K, unless stated otherwise. The effective dissipation was moderate with a relaxation time of 1 ps[34], allowing the thermostat-free region of the membrane to heat up well above $T_0$ with sufficiently high external excitation. All simulations were run for 0.2 ns and the averages of interest were calculated for the second half of the simulated time period, determined from the system energy time evolutions as sufficient to reach steady-state.

**Acknowledgment**

The authors are grateful to David T. Read, Jason P. Killgore, and Lauren Rast for valuable suggestions.

**References**

1.	Novoselov, K. S.; Geim, A. K.; Morozov, S. V.; Jiang, D.; Zhang, Y.; Dubonos, S. V.; Grigorieva, I. V.; Firsov, A. A., Electric Field Effect in Atomically Thin Carbon Films. *Science* 2004, 306, 666-669.
2.	Neto, A. H. C.; Guinea, F.; Peres, N. M. R.; Novoselov, K. S.; Geim, A. K., The electronic properties of graphene. *Reviews of Modern Physics* 2009, 81, 109-162.
3.	Novoselov, K. S.; Geim, A. K.; Morozov, S. V.; Jiang, D.; Katsnelson, M. I.; Grigorieva, I. V.; Dubonos, S. V.; Firsov, A. A., Two-dimensional gas of massless Dirac fermions in graphene. *Nature* 2005, 438, 197-200.
4.	Zhang, Y.; Tan, Y.-W.; Stormer, H. L.; Kim, P., Experimental observation of the quantum Hall effect and Berry's phase in graphene. *Nature* 2005, 438, 201-204.
5.	Novoselov, K. S.; Jiang, D.; Schedin, F.; Booth, T. J.; Khotkevich, V. V.; Morozov, S. V.; Geim, A. K., Two-dimensional atomic crystals. *Proc. Nat. Acad. Sci.* 2004, 102, 10451-10453.
6.	Meyer, J. C.; Geim, A. K.; Katsnelson, M. I.; Novoselov, K. S.; Booth, T. J.; Roth, S., The structure of suspended graphene sheets. *Nature (London)* 2007, 446, 60-63.
7.	Fasolino, A.; Los, J. H.; Katsnelson, M. I., Intrinsic ripples in graphene. *Nature Materials* 2007, 6, 858 - 861.
8.	Mermin, N. D., Crystalline Order in Two Dimensions. *Physical Review* 1968, 176, 250-254.
9.	Nelson, D.; Weinberg, S.; Piran, T., *Statistical Mechanics of Membranes and Surfaces*. 2 ed.; World Scientific: Singapore, 2004.
10.	Doussal, P. L.; Radzihovsky, L., Self-consistent theory of polymerized membranes. *Physical Review Letters* 1992, 69, 1209-1212.
11.	May, R. M., Quantum Statistics of Ideal Gases in Two Dimensions. *Physical Review* 1964, 135, A1515–A1518.
12.	Thompson-Flagg, R. C.; Moura, M. J. B.; Marder, M., Rippling of graphene. *Europhysics Letters* 2009, 85.
13.	Smolyanitsky, A.; Tewary, V. K., Atomistic simulation of a graphene-nanoribbon–metal interconnect. *Journal of Physics: Condensed Matter* 2011, 23.
14.	Zhu, W.; Low, T.; Perebeinos, V.; Bol, A. A.; Zhu, Y.; Yan, H.; Tersoff, J.; Avouris, P., Structure and electronic transport in graphene wrinkles. *Nano Letters* 2012, 12.
15.	Bao, W.; Miao, F.; Chen, Z.; Zhang, H.; Jang, W.; Dames, C.; Lau, C. N., Controlled ripple texturing of suspended graphene and ultrathin graphite membranes. *Nature Nanotechnology* 2009, 4, 562-566.
16.	Raux, P. S.; Reis, P. M.; Bush, J. W. M.; Clanet, C., Rolling Ribbons. *Physical Review Letters* 2010, 105.
17.	Bunch, J. S.; Zande, A. M. v. d.; Verbridge, S. S.; Frank, I. W.; Tanenbaum, D. M.; Parpia, J. M.; Craighead, H. G.; McEuen, P. L., Electromechanical Resonators from Graphene Sheets. *Science* 2007, 315, 490-493.

18. Klimov, N. N.; Jung, S.; Zhu, S.; Li, T.; Wright, C. A.; Solares, S. D.; Newell, D. B.; Zhitenev, N. B.; Stroscio, J. A., Electromechanical Properties of Graphene Drumheads. *Science* 2012, 336, 1557-1561.
19. Fogler, M. M.; Guinea, F.; Katsnelson, M. I., Pseudomagnetic Fields and Ballistic Transport in a Suspended Graphene Sheet. *Physical Review Letters* 2008, 101.
20. Tewary, V. K.; Yang, B., Singular behavior of the Debye-Waller factor of graphene. *Physical Review B* 2009, 79.
21. Iyakutti, K.; Surya, V. J.; Emelda, K.; Kawazoe, Y., Simulation of ripples in single layer graphene sheets and study of their vibrational and elastic properties. *Computational Materials Science* 2012, 51, 96-102.
22. Elmore, W. C.; Heald, M. A., *Physics of Waves*. Dover, Inc.: New York, 1985.
23. Pathria, R. K.; Beale, P. D., *Statistical Mechanics*. 3 ed.; Elsevier: 2011.
24. Maluckov, A.; Hadzievski, L.; Malomed, B. A., Fundamental solitons in discrete lattices with a delayed nonlinear response. *Chaos* 2010, 20.
25. Crassee, I.; Orlita, M.; Potemski, M.; Walter, A. L.; Ostler, M.; Seyller, T.; Gaponenko, I.; Chen, J.; Kuzmenko, A. B., Intrinsic Terahertz Plasmons and Magnetoplasmons in Large Scale Monolayer Graphene. *Nano Letters* 2012, 12, 2470–2474.
26. Pellegrino, F. M. D.; Angilella, G. G. N.; Pucci, R., Dynamical polarization of graphene under strain. *Physical Review B* 2010, 82.
27. Cole, R. M.; Mahajan, S.; Baumberg, J. J., Stretchable metal-elastomer nanovoids for tunable plasmons. *Applied Physics Letters* 2009, 95.
28. Sciammarella, C. A.; Lamberti, L.; Sciammarella, F. M.; Demelio, G. P.; Dicuonzo, A.; Boccaccio, A., Application of Plasmons to the Determination of Surface Profile and Contact Strain Distribution. *Strain* 2010, 46, 307–323.
29. Berber, S.; Kwon, Y.-K.; Tománek, D., Unusually High Thermal Conductivity of Carbon Nanotubes. *Physical Review Letters* 2000, 84, 4613–4616.
30. Evans, W. J.; Hu, L.; Keblinski, P., Thermal conductivity of graphene ribbons from equilibrium molecular dynamics: Effect of ribbon width, edge roughness, and hydrogen termination. *Applied Physics Letters* 2010, 96.
31. Lindsay, L.; Broido, D. A., Optimized Tersoff and Brenner empirical potential parameters for lattice dynamics and phonon thermal transport in carbon nanotubes and graphene. *Physical Review B* 2010, 81.
32. Smolyanitsky, A.; Killgore, J. P.; Tewary, V. K., Effect of elastic deformation on frictional properties of few-layer graphene. *Physical Review B* 2012, 85, 035412
33. Hoover, W. G., Canonical dynamics: Equilibrium phase-space distributions. *Physical Review A* 1985, 31, 1695–1697.
34. Hünenberger, P. H., *Thermostat Algorithms for Molecular Dynamics Simulations*. Springer Berlin / Heidelberg: 2005; Vol. 173, p 130.

**Figures**

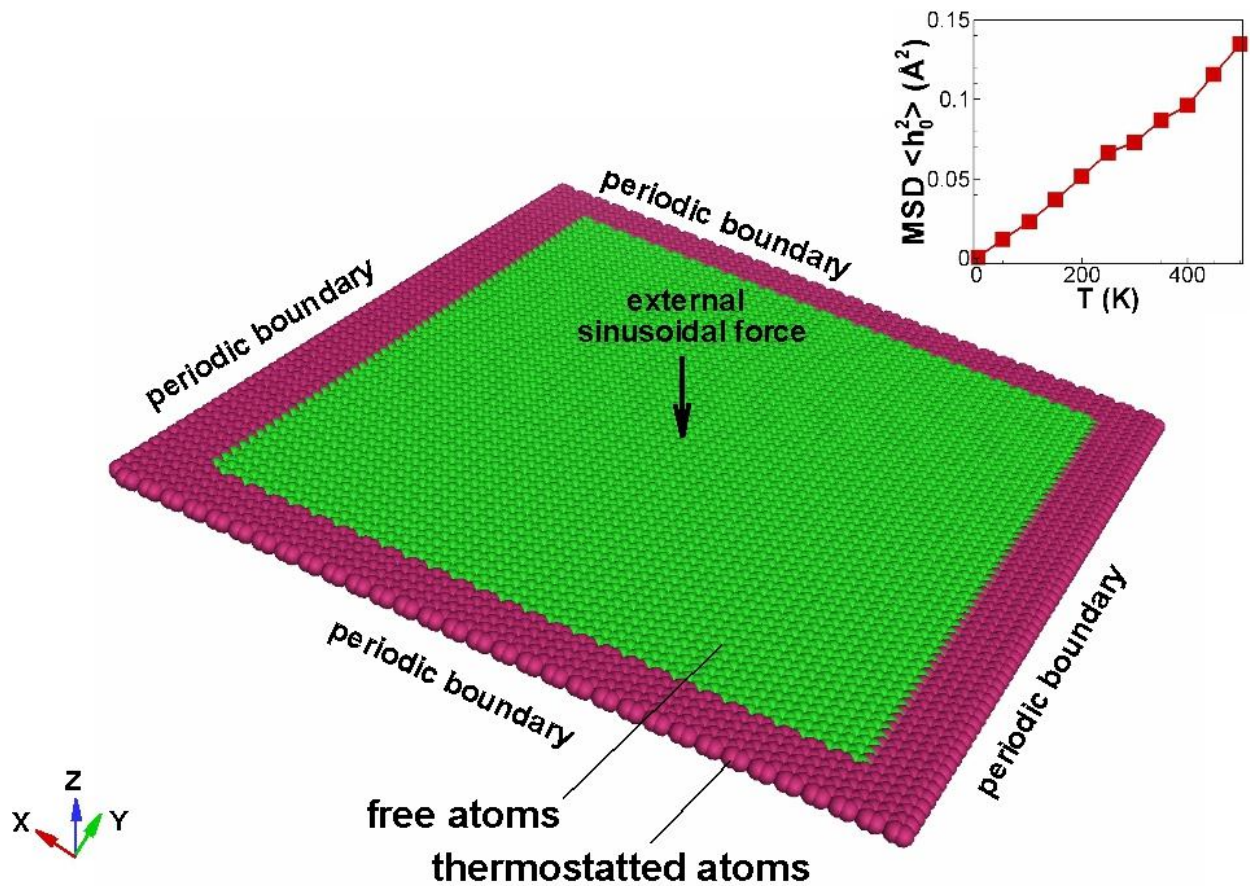

Figure 1. Simulation setup outlining the boundary conditions and thermally controlled regions. The inset shows linear dependence of the out-of-plane mean-square displacement $\langle h_0^2 \rangle$ on the membrane temperature in the case of no external excitation, as expected.

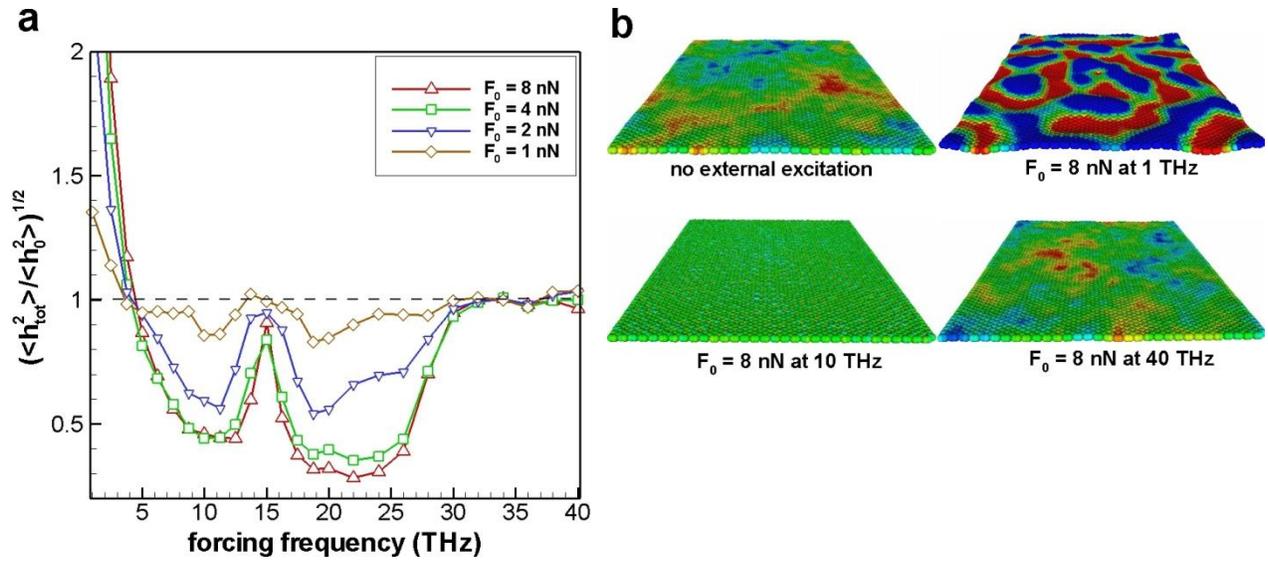

Figure 2. Out-of-plane root-mean-square ratio between externally excited and excitation-free systems $\varepsilon = \sqrt{\langle h_{tot}^2 \rangle / \langle h_0^2 \rangle}$ versus excitation frequency at various excitation amplitudes (a) and representative examples of local morphology at different excitation frequencies (b). The color range is constant from -0.8 Å (blue) to +0.8 Å (red).

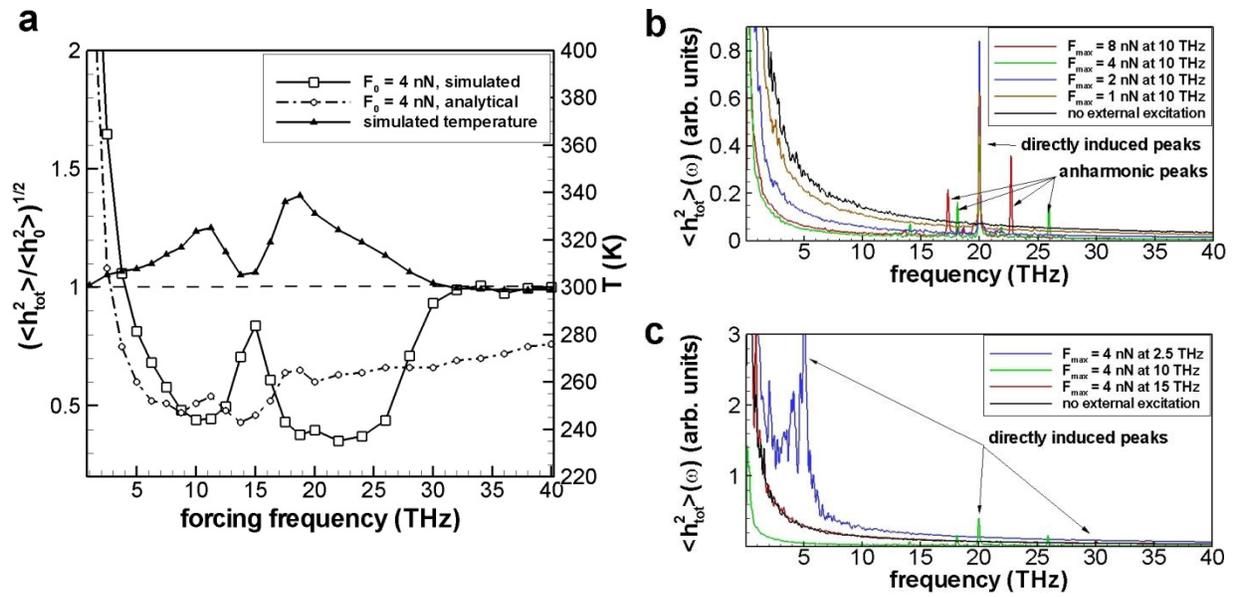

Figure 3. Simulated and analytically calculated normalized out-of-plane RMSD $\varepsilon$ (left Y-axis) and simulated temperature (right Y-axis) (a); Fourier transforms of $h_{tot}^2(t)$ at a constant excitation frequency, and various amplitudes (b), and constant excitation amplitude and various frequencies (c).

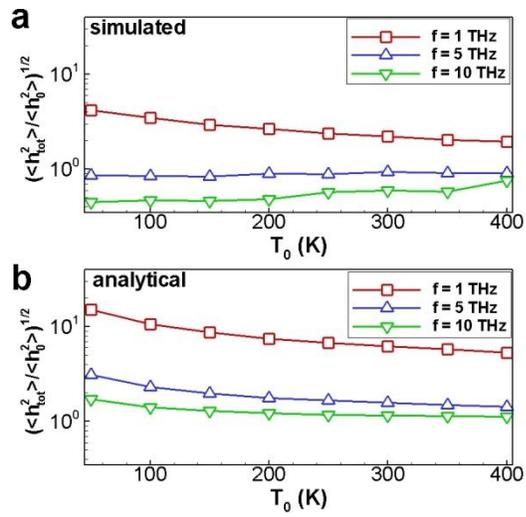

Figure 4. Simulated (a) and analytically calculated (b) values of $\varepsilon = \sqrt{\langle h_{tot}^2 \rangle / \langle h_0^2 \rangle}$ for various frequencies and constant $F_0 = 2\ nN$.

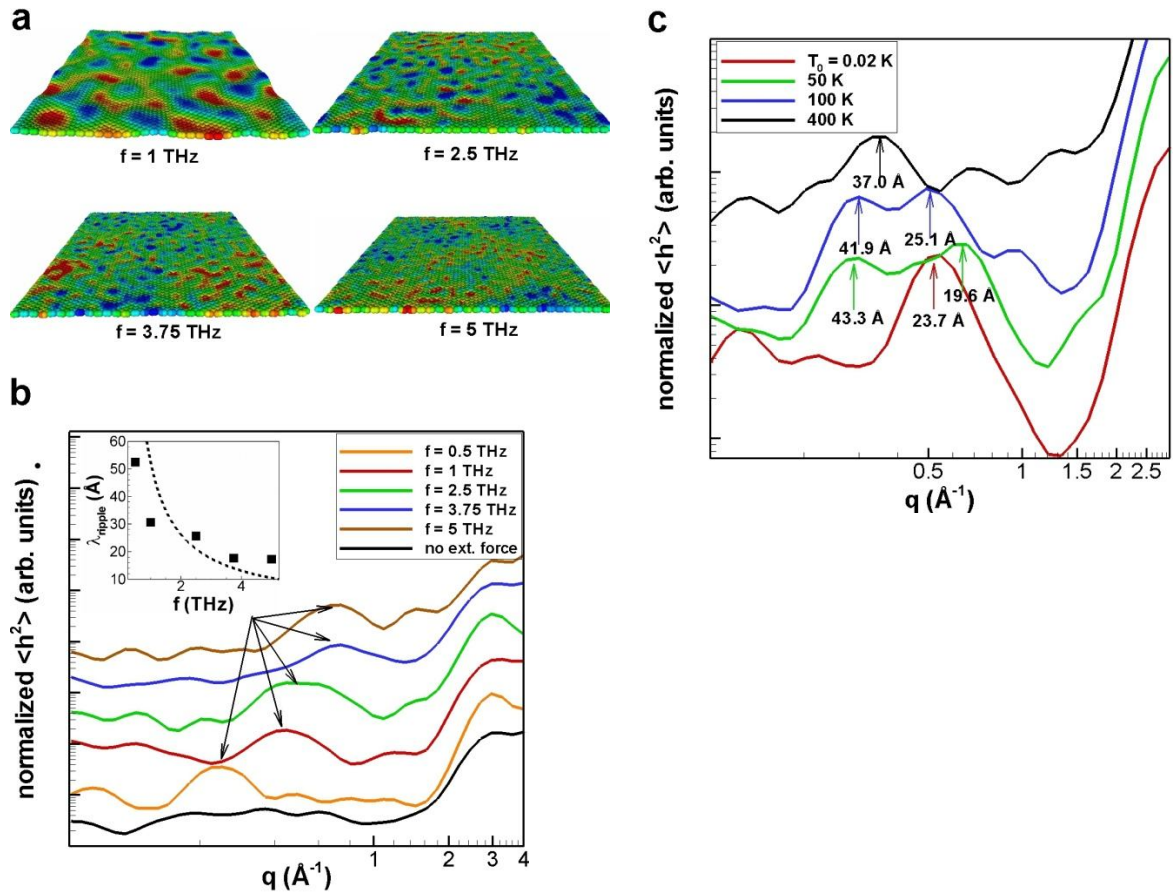

Figure 5. Simulated patterns for $F_0 = 4\ nN$ and different excitation frequencies (a), reciprocal space spectra of $h_{tot}^2$ at $F_0 = 4\ nN$ and $T_0 = 300$ K (b), and temperature-dependent reciprocal space spectra of $h_{tot}^2$ at $F_0 = 2\ nN$, at various temperatures (c).

**Supplementary Information for the paper:** *Manipulation of graphene's dynamic ripples by local harmonic out-of-plane excitation* **by A. Smolyanitsky and V.K. Tewary**

*1. Derivation of the out-of-plane MSD due to external excitation*

Consider a simple case of a rectangular membrane (dimensions $L_x \times L_y$) with clamped edges, externally excited, as shown in Fig 1 of main text, without the stochastic contribution (the membrane is flat initially). Although periodic boundaries were used in the simulations, we later show in the main text that an identical result of interest is obtained for a rectangular membrane regardless of the boundary condition. The solution for the external force $F(t) = F_0 \cos(\omega t)$ applied at $\left(\frac{L_x}{2}, \frac{L_y}{2}\right)$ is well-known and is a sum of the local modes[S1]:

$$h_{ext}(x, y, t) = \sum_{n_x, n_y} h_{0, n_x, n_y, ext}\left(\omega, \omega_{0, n_x, n_y}\right) \sin\left(\frac{n_x \pi x}{L_x}\right) \sin\left(\frac{n_y \pi y}{L_y}\right) \cos(\omega t + \varphi_{n_x, n_y}), \quad (S1)$$

where integers $n_x, n_y$ determine the corresponding resonant mode with

$\omega_{0, n_x, n_y} = c \sqrt{\left(\frac{n_x \pi}{L_x}\right)^2 + \left(\frac{n_y \pi}{L_y}\right)^2}$ ($c$ is the velocity of transverse waves in the membrane material); the amplitude $h_{0, n_x, n_y, ext}\left(\omega, \omega_{0, n_x, n_y}\right)$ is determined by the external source and includes proper Fourier "weighting", $\varphi_{n_x, n_y}$ is a phase shift due to damping. Although using Eq. (S1) in its general form to obtain $\langle h_{ext}^2 \rangle$ and $E_{ext}$ is straightforward, it is tedious and largely unnecessary for our purposes. Instead, we provide a simplified analysis to reveal the qualitative mechanism of the observed flattening. Let us consider a single-mode case $n_x = n_y = 1$, assuming this mode dominates the response of the membrane. The sum in Eq. (S1) is then replaced with a single term, in which

$$h_{0,1,1,ext}(\omega, \omega_0) = \frac{F_0}{m\sqrt{\gamma^2 \omega^2 + (\omega^2 - \omega_0^2)^2}}, \quad (S2)$$

where $\gamma$ is the effective damping constant as a result of dissipation on the externally induced vibrations, $\omega_0 = \omega_{0,1,1}$, and $m$ is the mass of a carbon atom. We obtain $\langle h_{ext}^2 \rangle$ by averaging over the period of excitation $\tau = 2\pi/\omega$, as well as the membrane surface $\Omega$:

$$\langle h_{ext}^2 \rangle = \frac{1}{L_x L_y \tau} \iint h_{ext}^2(x, y, t) d\Omega dt. \quad (S3)$$

Substituting the single-mode Eq. (S1) into Eq. (S3), with use of Eq. (S2), yields

$$\langle h_{ext}^2 \rangle = \frac{1}{4} \frac{F_0^2}{m^2(\gamma^2 \omega^2 + (\omega^2 - \omega_0^2)^2)} \quad (S4)$$

and $E_{ext} = \frac{m}{2} \langle h_{ext}^2 \rangle \omega^2$, thus determining all terms in Eq. (6) of the main text. Eq. (S4) is identical to Eq. (7) in the main text.

From the simulations, $c \cong 6000\ m/s$ and thus for $L_x = 13.6\ nm$ and $L_y = 15.8\ nm$ we obtain $f_0 = \omega_0/2\pi = 0.29\ THz$.

*2. Response to high-frequency excitation*

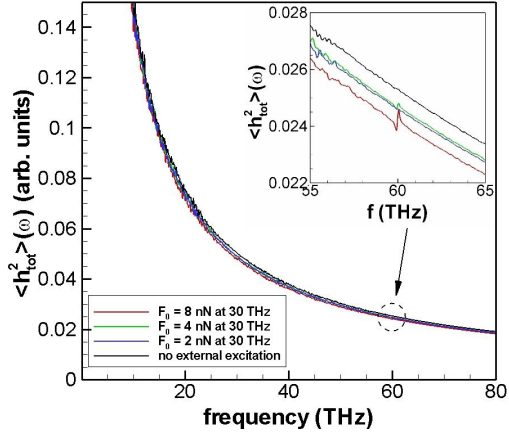

Figure. S1. Fourier transforms of $h_{tot}^2(t)$ at $f = 30\ THz$ and various external force amplitudes compared with the case of no external excitation.

*3. Low-frequency patterns for an edge-clamped circular membrane*

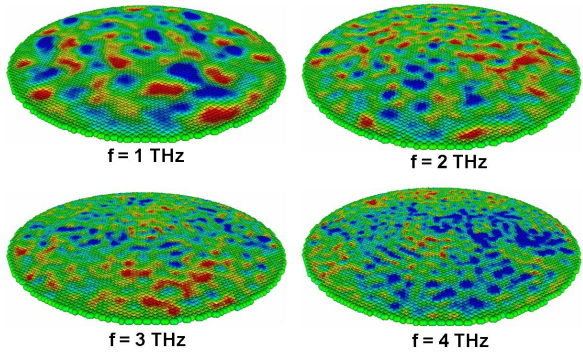

Figure S2. Simulated patterns for an edge-clamped circular (20 nm diameter) graphene membrane temperature-controlled ($T_0 = 300$ K) at the circumference at $F_0 = 4\ nN$ and different excitation frequencies. Color ranges adjusted appropriately due to rippling amplitudes decreasing with increasing excitation frequency.

*4. Normal-normal correlations at 300 K*

Fig. S4 shows the normal-normal correlations calculated for our system at 300 K with and without external input. The two peaks we calculated are within 3% and 14% from the Bragg peaks at 2.94 $Å^{-1}$ and 5.11 $Å^{-1}$, [S2] indicating that the Tersoff-Brenner potential we used is consistent with the LCBOPII potential used in the Monte Carlo simulations in [S2] for describing the bonding properties of graphene.

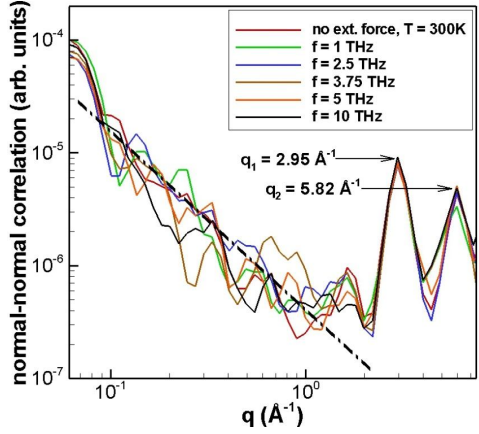

Figure S3. Normal-normal correlation as a function of the wave-vector magnitude, as defined in.[S3] The peaks determined by the structure of graphene are in good agreement with the data reported elsewhere.[S2]

*5. Excitation of rippling patterns by applying external force to a wider spot*

Shown in Figs. S4 (a,b) are some of the results obtained from applying external excitation to a ~ 1 nm –wide circular region (34 atoms total) centered at the intersection of the membrane's diagonals with a per-atom force of 0.2 nN. A rippling pattern in Fig. S4 (a) is clearly observed. The data in Fig. S4 (b) has been normalized and shifted along the Y-axis for clarity, similarly to Fig. 5 (b, c) of main text. Although the driven peak at f = 1 THz is close to that reported in the main text, it shifts less rapidly with increasing frequency. In addition, a stable peak at approximately half the size of the membrane is present regardless of the external excitation frequency. We attribute these observations to the driven spot size being only barely ten times smaller than the characteristic membrane size. We expect to observe the behavior reported in the main text when the spot size is significantly smaller than the membrane.

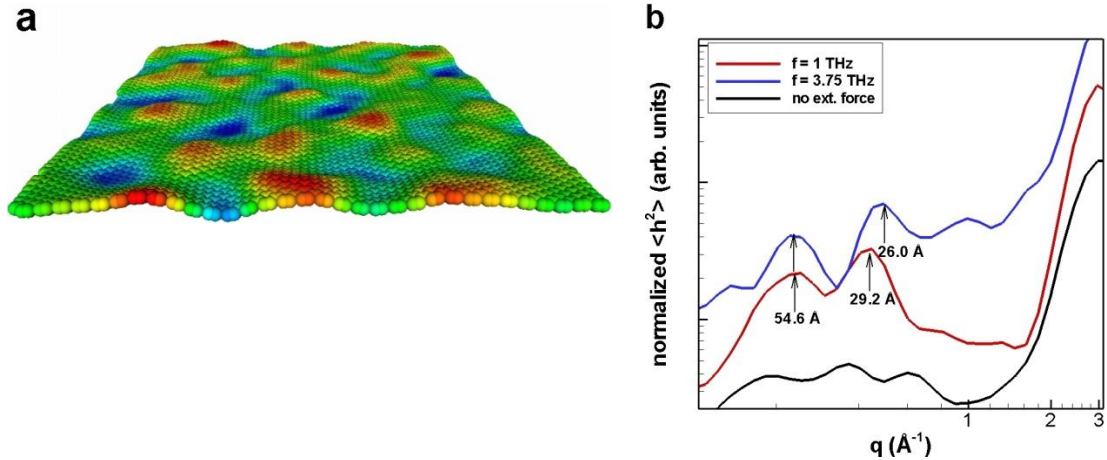

Figure S4. Simulated pattern at $T_0 = 300$ K, $F_0 = 34 \times 0.2\ nN = 6.8\ nN$, and f = 1 THz (a) and normalized reciprocal space distributions of $h_{tot}^2$ at f = 1 THz and 3.75 THz (b). The case without external excitation is provided as reference.

S1. Courant, R.; Hilbert, D., *Methods of Mathematical Physics, Vol. 1*. 3 ed.; Interscience Publishers: 1989.
S2. Fasolino, A.; Los, J. H.; Katsnelson, M. I. *Nature Materials* **2007,** 6, (11), 858 - 861.
S3. Nelson, D.; Weinberg, S.; Piran, T., *Statistical Mechanics of Membranes and Surfaces*. 2 ed.; World Scientific: Singapore, 2004.